\documentclass[12pt,epsf]{article}

\def\be{\begin{equation}}
\def\ee{\end{equation}}
\def\bea{\begin{eqnarray}}
\def\eea{\end{eqnarray}}

\def\be{\begin{equation}}
\def\ee{\end{equation}}
\def\bea{\begin{eqnarray}}
\def\eea{\end{eqnarray}}

\def\case#1/#2{\textstyle\frac{#1}{#2}}
\def\k0{\kappa_{0}}

\begin{document}
\begin{titlepage}

\vspace{.7in}

\begin{center}
\Large
{\bf Self-gravitating Bjorken Flow }\\
\vspace{.7in}
\normalsize
\large{ 
 $Alexander Feinstein$
}\\
\normalsize
\vspace{.4in}

{\em Dpto. de F\'{\i}sica Te\'orica, Universidad del Pa\'{\i}s Vasco, UPV/EHU \\
Apdo. 644, E-48080, Bilbao, Spain}\\
\vspace{.2in}
\end{center}
\vspace{.3in}
\baselineskip=24pt

\begin{abstract}
\noindent I present a solution to the full Einstein-fluid equations representing a self-gravitating Bjorken flow. The motion and the geometry become inhomogeneous in the plane transversal to the  flow and the energy density profile acquires, due to gravity, corrections in terms of proper time as compared to the original test hydrodynamics. The transverse distribution of energy density, for example, becomes $\epsilon(\tau,r)/\epsilon(\tau,0)\,=\, \cosh^{-4}{\left(3ar\right)}$.
\end{abstract}

\vspace{.3in}

\end{titlepage}

Bjorken  flow \cite{Bjorken} represents the most fascinating  application  of relativistic hydrodynamics to an extremely complex physical system describing an average motion of partons resulting in a collision of heavy ions. The application of hydrodynamics to similar problems was pioneered by Landau \cite{Landau} to describe the high-energy multiparticle collisions.
Both in Bjorken and Landau descriptions it is assumed  that after the collision of heavy ions the mean free path of the constituencies is  short enough, so that the hydrodynamical description is meaningful. The difference between the two pictures is  in the symmetry assumptions. In Bjorken hydrodynamics, one assumes the so-called boost invariance, so that the energy density only depends on proper time $\tau$, while in Landau picture, no such symmetry restriction is made and the density may be a function of all spatial coordinates. Needless to say that Bjorken flow is a particular and simpler  version of Landau hydrodynamics, nevertheless,  it is  surprising  it works so good \cite{experiments}.    A different  and renewed motivation   in these studies comes from their relation to  the   AdS-CFT correspondence conjecture  \cite{maldacena},  \cite{Son}, \cite{JanikPeschanski},   because  they serve as an input to understand the highly nontrivial behavior of Quantum Chromodynamics in a strong coupling regime.

The symmetries one imposes on the Bjorken flow  are:
 the  boost symmetry along the beam,  translational and the rotational invariance in the transverse plane. These symmetries allow one to parametrise the ``future wedge" of the Minkowski spacetime in the following way

\be
ds^2=\,-d\tau^{2} \,+\,\tau^{2} \, d\eta^{2}\, + \,d\rho^{2}\,+\,\rho^{2} d\phi^2 \label{metric1} 
\ee

Here $\tau$ is the proper time  and $\eta$ is historically called the rapidity and the rest are usual cylindrical coordinates. 

For further purposes  I will write the metric in the following form

\be
ds^2=\,-d\tau^{2} \,+ \,d\rho^{2}\,+\,\tau\rho \,\left(\tau/\rho \, d\eta^{2}\, +\,\rho/\tau\, d\phi^{2}\right) \label{metric2} 
\ee

The spacelike Killing fields $\frac{\partial}{\partial{\phi}}$ and  $\frac{\partial}{\partial{\eta}}$ are  the rotational and the boost Killing vectors respectively, and the form of the line element (\ref{metric2}) gives one an idea as to how to proceed in order to generalise this set up to a General Relativistic flow. 
 
Now, if one introduces an ideal fluid with the linear equation of state $p=1/3\, \epsilon$ with the energy density and the pressure such that these only depend on the proper time $\tau$, in other words the fluid velocity has no tilt, it is quite easy to solve the hydrodynamical equations and to obtain that the fluid density scales as $\propto\tau^{-4/3}$. This is quite close to the experimental picture, \emph{grosso modo}, within the range where the hydrodynamics makes sense. 

On the other hand, the above picture is quite idealised, of course, some amount of viscosity should be added and some symmetries relaxed \cite{gubser}, nevertheless, it is quite surprising that the picture works so well. 

While the Bjorken hydrodynamics deals exclusively with the flow on a given  fixed flat background geometry (test hydrodynamics), and given that even slight relaxations of symmetry would leed to quite complicated nonlinear hydrodynamical equations, even the so-called Khalatnikov solution \cite{Khalatnikov} of a 1+1 dimensional flow is quite a mess \cite{Peschanski}, it is yet another pleasant surprise that one may integrate an exact general relativistic solution which describes self gravitating Bjorken flow. The main purpose of this Letter is to present such a solution and to compare it to the original Bjorken's test hydrodynamics. As a by-product, I will also obtain solutions to the test hydrodynamics without tilt on a class of cylindrical geometries.

I will stick to the original Bjorken picture as close as possible and will assume that the fluid velocity has no tilt (see, however, a comment after the equation (\ref{equation of state}).) Nevertheless, since one must solve selfconsitently the coupled Einstein-fluid equations, one can not expect that the geometry would share all the above mentioned symmetries. This is the essence of the Einstein theory, the matter influences the geometry which in turn changes its motion.  We may keep the boost and the rotational Killing vectors   intact, however, there is no reason why the line element should not depend both on $\tau$ and $\rho$ coordinates. In fact, the form of the line element (\ref{metric2}) indeed suggests the dependence on $\rho$.  I will therefore assume the following geometry 
\be
ds^2=\,f\left(\,-dt^{2} \,+ \,dr^{2}\,\right)+\,g\,\left( q \, d\eta^{2}\, +\, q^{-1} \, d\phi^{2}\right) \label{EinsteinRosen} 
\ee 

Here $f$, $g$ and $q$ are functions of both  $t$ and $r$ the ``conformal" coordinates which are  labled differently  to distinguish them from the proper time and the proper distance coordinates. Obviously for the original Bjorken flow these functions are $f=1$, $g=t\,r$ and $q=\,t/r$.  The equation (\ref{EinsteinRosen}) seems a natural generalization of Bjorken geometry, while the attempts to use the homogeneous Kasner, or flat FRW line element, as some authors do, fails to address the inherited symmetries of the problem.

I now specify the matter. The perfect fluid is assumed to have a linear equation of state, and because the fluid flow is irrotational, which is a must in this geometry, one may introduce the following velocity potential $\sigma$ \cite{schutz}, \cite{Liang}:
 \be
u^{\mu}= \sigma^{\mu}/ \sqrt{-\sigma_{\alpha}\sigma^{\alpha}}
\label{velocity potential} 
\ee

 As will bee  later seen,  the velocity potential is an extremely useful tool to solve the hydrodynamics.

Having done so,  one may further define the kinetic scalar (``enthalpy") $X=-1/2\, \sigma_{\alpha}\sigma^{\alpha}$. The pressure and the energy density  can then be expressed as follows \cite{Diez-TejedorFeinstein}
 \be
 p=p(X), \, \epsilon= 2X\,p'-p
  \ee
  Here prime, as usual, stands for the derivative of the function with respect to its argument.
  
  If the equation of state is linear $p=w\,\epsilon$, one may further write \cite{Liang}, \cite{Diez-TejedorFeinstein} 
  
  \be
 p(X)=X^{\left(w+1\right)/2w}
 \label{equation of state} 
  \ee

  We now assume that we have chosen our coordinates comoving with the fluid flow as in the original setting so that the velocity has only a zero component $u_{0}$. This translates, in terms of the velocity potential, that $\sigma$  is a function of $t$ alone. In fact, if one would  have even allowed a tilt, so that the velocity would ``catch"  a component in the transversal direction ($\frac{\partial \sigma}{\partial r}\neq\,0$), as for example in (\cite{gubser}), there would still be  a way to   introduce a new coordinate system comoving with the fluid, and maintain the form of the metric \cite{wainwright}. This would not necessarily be true for the test hydrodynamics, and  it may also spoil the separability of the metric functions which I will assume in the future. 
  
  The full Einstein equations for the line element (\ref{EinsteinRosen}) with the fluid specified above are found in \cite{Liang}. It is instructive, however, to  display the dynamical equation for the velocity potential  $\nabla_{\mu}\left(p'\,\sigma^{\mu}\right)\,=0$ \cite{Diez-TejedorFeinstein}. This  reads:
  \be
 \frac{1}{v_{s}^2}\,\ddot{\sigma}+\left[\dot{g}/g-\,\frac{1}{2}\left(\frac{1}{v_{s}^2} -1\right)                                                       \,\dot{f}/f\right]\,\dot{\sigma}\,=\,0,
 \label{dynamics} 
\ee

where the velocity of sound $v_{s}$ is given by
\be
v_{s}^2\,=\, \frac{p'\left(X\right)}{2Xp''\left(X\right)+p'\left(X\right)},
\ee
 and can be easily obtained from the relations (5) and the expression $v_{s}^2\,= \frac{\partial p}{\partial \rho}$. 
  
  Note that the transversal degree of freedom of the metric $q$ plays no explicit role in the dynamics of the fluid, but does influence the flow via the  full Einstein equations  contributing to the longitudinal expansion $f$. Both, though, the longitudinal expansion $f$ and the function $g$, which is proportional to the area of the isometry group orbits, do appear in the equation. These functions ($p$, $g$ and $f$) are determined by the Einstein Equations.
  
   Using the  linear equation of state (\ref{equation of state}), so that $v_{s}^{2}=w$, and assuming that all the functions of the metric are separable, $f\,=\,f_{T}\,f_{R}$, $g\,=\,g_{T}\,g_{R}$ and so on,   one may easily integrate the dynamical equation (\ref{dynamics}) to get
  \be
 \dot{\sigma}= b\, \frac{f_{T}^{(1-w)/2}}{g_{T}^{w}}, \label{solutiondynamics}
 \ee

 where lower index $T$ indicates the time dependent part of the respective function and $b$ is an arbitrary integration constant. The solution (\ref{solutiondynamics}) represents therefore Bjorken flow on a generalised geometry given by the line element (\ref{EinsteinRosen}). Hence, given the background geometry (\ref{EinsteinRosen}) with separable functions, the velocity potential is given by the solution (\ref{solutiondynamics}), and therefore all the kinematical fluid variables, may be easily evaluated.
 For the Bjorken geometry, and $w=1/3$ we get $\sigma\propto\,t^{2/3}$ which leaves us with $\epsilon \propto\,t^{-4/3}$ and $t$ is a proper time coordinate in this case. For the spatially flat FRW geometry sourced by the same fluid (radiation) $f \propto\, g\propto\, t^2$, we get $\sigma \propto\,t$, $X\propto\, t^{-2}$ and $\epsilon \propto\, t^{-4}$ which becomes $\epsilon \propto\, \tau^{-2}$ in terms of proper time. One may play round with the equation (\ref{dynamics}) further, but I will leave it for elsewhere.
 
 The equation (\ref{dynamics}), though important, is only a part of the full Einstein Equations and describes  a test hydrodynamics on a given geometry. If one solves the rest of the equations with the following initial and boundary conditions compatible with the Bjorken geometry, $g\,\rightarrow\ \,t \,r\,$ as $t$ and $r$ $\rightarrow\,0$, and $t\,\rightarrow\,0$ as $\tau\,\rightarrow\,0$  one can integrate the following  solution to the Einstein Equations:
 
\begin{eqnarray}
f &=& \sinh^{4}{\left(at\right)}\cosh^{2}{\left(3ar\right)} \nonumber 
\\ g &=& \sinh{\left(at\right)}\sinh{\left(3ar\right)}\cosh^{-2/3}{\left(3ar\right)} \nonumber
\\ q &=&\sinh^{3}{\left(at\right)}\sinh{\left(3ar\right)} 
\label{solution}
\end{eqnarray}

The velocity potential for this solution is given by
\be 
\sigma=(15\,a^{2})^{1/4}\cosh{\left(at\right)}
\ee

while the energy density becomes

\be
\epsilon= 15a^{2} \sinh^{-4}{\left(at\right)}\cosh^{-4}{\left(3ar\right)}
\ee
Here $a$ is a free parameter specifying the density.
In fact this solution was derived some years ago by the author together with J. Senovilla \cite{FeinsteinSenovilla} in the context of inhomogeneous cosmology, but it could not occur to us then that the solution describes the inhomogeneous  selfgravitating Bjorken flow. 
Indeed, since the coordinate $t$ is not the proper time, we find that the coordinate time  $t \propto \tau^{1/3}$  and therefore the energy density scales as 
\be
\epsilon  \propto \tau^{-4/3},
\ee

to the lowest order in proper time. However, some corrections are due.  Because  we have chosen our coordinates in a way that $r$ remains constant along the fluid lines, the proper time is given by the following expression
\be
\tau\,=\, \int^{t}_{0}\sqrt{f}dt\,= \frac{1}{3a}\,(at)^{3}+ \frac{1}{15a}\,(at)^{5}+ \mathcal{O}((at)^{7})
\ee
The energy density $\epsilon$ then evolves as (just the first two terms)
\be
\epsilon\, \propto\, \alpha\left(a\right) \,\tau^{-4/3} -\, \beta\left(a\right)\,\tau^{-2/3}
\label{density},
\ee
here $\alpha$ and $\beta$ are both positive functions of the parameter $a$, I have assumed a constant $r$, and have used the lowest order of the proper time expansion.
Of course, having at hand the exact solution, there is no need in the series expansions, nevertheless, these are instructive in order to further elucidate the physics. As one can easily see from the equation (\ref{density}), the second term becomes dominating at late times, where the energy density appears to become negative. This is an \emph{artifact} of the  series expansions, for the exact energy density never becomes negative. On the other hand, while the first term scales as the energy density of the ideal fluid, the second term acts  as if it were viscosity by enhancing the fall off of the evolving density. Of course, the fluid remains  inviscid for all the time, and this effect is purely due to self-gravity. Assuming the Stefan-Boltzmann's law ($\epsilon \, \sim \, T^{4}$) one can easily find the temperature distribution and then define the temperature contrast $\delta T$ as
\be
\delta T \,=\, \frac{T-T_{B}}{T_{B}},
\ee

 where $T$ is the temperature found from the exact solution, while $T_{B}$ represents the temperature of the test flow. This temperature contrast evolves as $\delta T \,  \sim \,- \tau^{1/3}$. Another interesting physical quantity is the  distribution of energy density in the transverse coordinate 
\be 
 \epsilon(\tau,r)/\epsilon(\tau,0)\,=\,\cosh^{-4}{\left(3ar\right)},
\ee
which is given by a neat simple expression in terms of ``conformal" distance $r$. It is assumed that this quantity is proportional to the distribution of nucleous in the fireball, the actual numbers, of course, would depend on the colliding constituencies.

To close, I have presented an exact solution to the Einstein-fluid equations which describes  selfgravitating Bjorken flow. The gravity  changes the energy density distribution and its evolution in proper time. From the exact expressions one may easily find all relevant kinematical and thermodynamical quantities. As a by-product, I have obtained some test hydrodynamical solutions  on the expanding cylindrical geometry. It would be interesting in the future to study more the test hydrodynamics with nonlinear equations of state, as well as to consider the solution (\ref{solution}) as an input to study the ADS/CFT correspondence in situations where the geometry is both inhomogeneous and evolving in time.

\centerline{\bf Acknowledgments}
It is a pleasure to thank Manuel Valle and Ra\"ul Vera  for valuable comments.
This work is partially supported by the Basque government Grant GICO7/51-IT-221-07,
the Spanish Science Ministry Grant FIS2010-15492 and the UFI 11/55 program of the University of the Basque country UPV/EHU.

\end{document}